\documentclass[conference]{IEEEtran}
\usepackage{amssymb} % for drawing double-barred R, Z, C, etc.

\begin{document}
\input{epsf}

%\title{The Schr\"{o}dinger Equation as Iterative Soft-Decision Optimization Algorithm in a Continuous-time Version}
\title{Deriving Schr\"{o}dinger Equation From \\A Soft-Decision Iterative Decoding Algorithm}

\author{
\authorblockN{Xiaofei Huang\authorrefmark{1}, Xiaowu Huang\authorrefmark{2}} \\
\authorblockA{\authorrefmark{1}School of Information Science and Technology\\
Tsinghua University, Beijing, P.~R.~China, 100084 \\
Email: huangxiaofei@ieee.org\\}
\authorblockA{\authorrefmark{2}Department of Physics,
Anhui University, Hefei, Anhui Province, P.~R.~China, 230039}}

% avoiding spaces at the end of the author lines is not a problem with
% conference papers because we don't use \thanks or \IEEEmembership
%Homer Simpson\authorrefmark{2},
%Montgomery Scott\authorrefmark{3} and
%Eldon Tyrell\authorrefmark{4}}
%Telephone: (800) 555--1212, Fax: (888) 555--1212}
%\authorblockA{\authorrefmark{4}Tyrell Inc., 123 Replicant Street, Los Angeles, California 90210--4321}}

\maketitle

\begin{abstract}
The belief propagation algorithm has been recognized in the information theory community 
	as a soft-decision iterative decoding algorithm.
It is the most powerful algorithm found so far for attacking hard optimization problems in channel decoding. 
Quantum mechanics is the foundation of modern physics
	with the time-independent Schr\"{o}dinger equation being one of the most important equations.
This paper shows that the equation can be derived from a generalized belief propagation algorithm.
Such a connection on a mathematical basis might shed new insights into the foundations of quantum mechanics and quantum computing.
\end{abstract}

\def\thefootnote{\fnsymbol{footnote}}
\setcounter{footnote}{-1}
\footnote{An extended abstract version of this paper has been presented at
       	International Conference on Quantum Foundation and Technology, August, 2006, HangZhou, China (ICQFT'06).}

%------------------------------------------------------------
\section{Introduction}
%------------------------------------------------------------

The maximum likelihood (ML) decoding of channel codes
	can be viewed as a computational task of finding the global minimum of an energy (objective) function.
The belief propagation (BP) algorithm~\cite{Pearl88} 
	is the most powerful one recognized by the information theory community~\cite{Mackay99}
	to accomplish the task.
The new generation channel codes such as Turbo codes and low-density parity-check (LDPC) codes
	combined with BP decoding can achieve remarkable performance close to the Shannon limit.
	
The {\it a posteriori} probability (APP) algorithm~\cite{Fossorier99} is a simplified variation of the BP algorithm.
Given a multivariate energy function $E(x_1, x_2, \ldots, x_n)$ of the following form
\[ E(x_1, x_2, \ldots, x_n) = \sum_i \left (e_i (x_i) + \sum_{j, j < i} e_{ij} (x_i, x_j) \right) \ , \]
with the assumption of the symmetric binary component functions $e_{ij}(x_i, x_j)$, 
	i.e., $e_{ij}(x_i, x_j) = e_{ji}(x_j, x_i)$, for any $i, j$.
The APP algorithm can be applied to find an approximate solution to minimize the energy function. 
It is based on a method of updating and passing $n$ messages, $\psi_i(x_i, t)$ for $i=1,2,\ldots,n$, 
	in an iterative way as follows,
\begin{eqnarray}
\lefteqn{ \psi_i(x_i, t+1) = \frac{1}{Z_i(t+1)} e^{-e_i(x_i) /\hbar} \cdot } \nonumber \\
& & \prod_{j \not = i} \left( \sum_{x_j} e^{- e_{ij}(x_i, x_j) /\hbar} \psi_j(x_j, t) \right), 
\label{app_algorithm}
\end{eqnarray}
where $\hbar$ is a positive constant, related to the channel characteristics in channel decoding.
$Z_i(t+1)$ is a normalization factor at time $t+1$, such that
\[ \sum_{x_i} \psi_i(x_i, t+1) = 1, \quad \mbox{for $i=1,2,\ldots, n$} \ . \]

The message $\psi_i(x_i, t)$ is a soft-decision for assigning variable $x_i$ at time $t$. 
It is a real-valued, non-negative function called the soft-assignment function in this paper.
It measures in a quantitative way the preferences over different values of $x_i$
	for minimizing the energy function.
The best candidate value for assigning $x_i$ at time $t$ is the one of the highest function value $\psi_i(x_i, t)$.
Often times at decoding channel codes, 
	each soft assignment function $\psi_i(x_i, t)$ 
	is progressively peaked at one variable value while the rest 
	reduced to zero as the iteration proceeds.
That is to say that the algorithm eventually decides on a unique value for each variable at those instances.
The density evolution~\cite{Richardson01} is a powerful technique 
	invented by the information theory community 
	to understand and analyze this kind of processes.
%------------------------------------------------------------
\section{A Generalization of the APP Algorithm}
%------------------------------------------------------------

The difference equations~(\ref{app_algorithm}) of the APP algorithm,
	can be generalized by raising the soft-assignment function $\psi_i(x_i, t)$ 
	at the right side of the equations to a power $\alpha$,
\begin{eqnarray}
\lefteqn{\psi_i(x_i, t+1) = \frac{1}{Z_i(t+1)} e^{-e_i(x_i) /\hbar} \cdot } \nonumber \\
&& \prod_{j \not = i} \left( \sum_{x_j} e^{- e_{ij}(x_i, x_j) /\hbar} |\psi_j(x_j, t)|^{\alpha} \right). 
\label{app_algorithm_a}
\end{eqnarray}
When $\alpha = 1$, the above generalization falls back to the original one.
	
It has been shown that the BP algorithm can only converge to a fixed point 
	that is also a stationary point of the Bethe approximation to the free energy~\cite{Yedidia05}. 
It is also not hard to prove that each valid codeword can be a fixed point  
	when the APP algorithm is applied to decode a LDPC code
	with the degree of each variable node $d_v \ge 2$ (or the BP algorithm when $d_v \ge 3$).
The algorithm will converge with an exponential rate to a fixed point of this kind 
	when it evolves into a state close enough to any one of them.

To improve the performance of the APP algorithm further,
	we can smooth the soft assignment functions $\psi_i(x_i, t)$ to
	prevent the algorithm from being trapped to an un-desired fixed point.
One way to smooth the soft assignment function $\psi_i(x_i, t)$ is given as follows,
\[ \psi^{'}_i(x_i, t) = (1-\beta) \psi_i(x_i, t) + \beta / |D_i| \ , \]
where $|D_i|$ is the domain size of variable $x_i$, 
	and the parameter $\beta$ is the smoothing factor satisfying $0 \le \beta \le 1$.
When $\beta = 1$, the function $\psi_i(x_i, t)$ is completely smoothed out.

The smoothing operation defines an operator on the soft assignment functions $\psi^{'}_i(x_i, t)$, 
	denoted as $\cal S(\cdot)$.
With that definition, we can generalize the APP algorithm~(\ref{app_algorithm_a}) further as follows,
\begin{eqnarray}
\lefteqn{ \psi_i(x_i, t+1) = \frac{1}{Z_i(t+1)} {\cal S} {\Big (} e^{-e_i(x_i) /\hbar} \cdot } \nonumber \\
	&& \prod_{j \not = i} ( \sum_{x_j} e^{- e_{ij}(x_i, x_j) /\hbar} |\psi_j(x_j, t)|^{\alpha}) {\Big )}. 
\label{app_algorithm_b}
\end{eqnarray}

It has been found that the generalized APP algorithm~(\ref{app_algorithm_b}) can sometimes significantly improve 
	the performance of the original one at decoding LDPC codes.
We have observed improvements over $1dB$ to $2dB$
	in our experiments at decoding commercial irregular LDPC codes 
	(such as LDPC codes used for China's HDTV) and regular experimental LDPC codes.

%%%%%%%%%%%%%%%%%%%%%%%%%%%%%%%%%%%%%%%%%%%%%%%%%%%%%%%%%%%%%%%%%%%%%%%%%%%%%%%%%%%%%%%%%%%%%%%%%%%%%
\section{Deriving Schr\"{o}dinger Equation}
%%%%%%%%%%%%%%%%%%%%%%%%%%%%%%%%%%%%%%%%%%%%%%%%%%%%%%%%%%%%%%%%%%%%%%%%%%%%%%%%%%%%%%%%%%%%%%%%%%%%%

Let the parameter $\alpha = 2$ in the generalized AP algorithm~(\ref{app_algorithm_b}).
If all variables $x_i$s are in a continuous domain, the generalized APP algorithm~(\ref{app_algorithm_b}) becomes
\begin{eqnarray}
\lefteqn{\psi_i(x_i, t+1)= \frac{1}{Z_i(t+1)} {\cal S} {\Big (}  e^{-e_i(x_i) /\hbar} \cdot } \nonumber \\
&& \prod_{j, j \not = i} \int d x_j~e^{-e_{ij}(x_i, x_j) /\hbar} |\psi_j (x_j, t)|^2 {\Big )} .
\label{app_algorithm_c}
\end{eqnarray}

The soft assignment function $\psi_i(x_i, t)$ can be generalized from a real-valued, non-negative function
	to a function over the complex domain $\mathbb{C}$.
It has no impact on the optimization power of the generalized APP algorithm~(\ref{app_algorithm_c}).
In this case, it is the magnitude of the function $|\psi_i(x_i, t)|$ instead of itself
	that measures the preferences over different values of $x_i$.
For $\psi_i(x_i, t) \in \mathbb{C}$, $|\psi_i(x_i, t)|$ is defined as $\sqrt{\psi^{*}_i(x_i, t) \psi_i(x_i, t)}$.

Let $\Delta t$ be an infinitesimal positive value 
	and the soft assignment function at $t + \Delta t$  be $\psi_i (x_i, t + \Delta t)$.
The difference equations~(\ref{app_algorithm_c}) of the generalized APP algorithm
	in a continuous time version is
	
\begin{eqnarray}
\lefteqn{ \psi_i(x_i, t + \Delta t) = \frac{1}{Z_i (t + \Delta t)} {\cal S} {\Big (} \psi_i(x_i, t) e^{-(\Delta t/\hbar)e_i(x_i)} \cdot } \nonumber \\
&& \prod_{j, j \not = i} \int d x_j~e^{-(\Delta t/\hbar)e_{ij}(x_i, x_j)} |\psi_j (x_j, t)|^2 {\Big )}. 
\label{update_rule5}
\end{eqnarray}
When $\Delta t \rightarrow 0$, the term inside the operator ${\cal S}(\cdot)$ 
	at the right side of (\ref{update_rule5}) approaches $\psi_i(x_i, t)$.

Starting from an initial state,
	the generalized APP algorithm described by (\ref{update_rule5}) 
	will evolve toward one of its equilibriums over time.
It will be shown in the following that 
	the algorithm at its equilibrium is, in fact, the time-independent Schr\"{o}dinger equation.

Since variable $x_i$ is in a continuous domain,
	let the smoothing operator ${\cal S}(\cdot)$ on the soft assignment function $\psi_i(x_i, t)$ be defined as
\[ {\cal S}(\psi_i(x_i, t)) = \int K(u - x_i) \psi_i(u, t) ~d u \ , \]
	where $K(x)$ is a smoothing kernel.
If $x_i$ is in the one dimensional space $\mathbb{R}$, 
	we can choose the following Gaussian function as the smoothing kernel $K(x)$,
\begin{equation}
K(x) = \frac{1}{\sqrt{2 \pi \Delta t} \sigma_i} e^{ - x^2 / 2 \sigma^2_i \Delta t} \ . 
\label{gaussian_kernel_1d}
\end{equation}

With the Gaussian smoothing kernel, the dynamic equations~(\ref{update_rule5}) become
\begin{eqnarray}
\lefteqn{ \psi_i(x_i, t + \Delta t) = \frac{1}{Z_i (t + \Delta t)} \cdot } \nonumber \\
	 && \int d u~\frac{1}{\sqrt{2 \pi \Delta t} \sigma_i} e^{ - (u - x_i)^2 / 2 \sigma^2_i \Delta t} \psi_i(u, t) e^{-(\Delta t /\hbar) e_i(u)} \cdot \nonumber \\
	 && \prod_{j, j \not = i} \int d x_j~e^{-(\Delta t /\hbar)e_{ij}(u, x_j)} |\psi_j (x_j, t)|^2  \ . 
\label{update_rule10}
\end{eqnarray}
Expanding the right side of the above equation into a Taylor series with respect to $\Delta t$ 
	and let $\Delta t \rightarrow 0$,
	we have 
\begin{eqnarray}
\lefteqn{ \frac{\partial \psi_i (x, t)}{\partial t}=  \frac{\sigma^2_i}{2}\frac{\partial^2 \psi (x_i, t)}{\partial x^2_i} - } \nonumber \\
&& V_i (x_i)\frac{1}{\hbar} \psi_i(x_i, t) + \varepsilon_i (t) \psi_i(x_i, t) \ , 
\label{update_rule11a}
\end{eqnarray}
where 
\[ V_i(x_i) = e_i(x_i) + \sum_{j, j \not = i} \int d x_j~e_{ij}(x_i, x_j) |\psi_j (x_j, t)|^2 \ , \]
and
\[ \varepsilon_i (t) = -\frac{d~Z_i(t)/d~t}{Z^2_i(t)} \ . \]

Let the operator $\nabla^2_i$ be defined as
\[ \nabla^2_i \psi (x_i, t) = \frac{\partial^2 \psi (x_i, t)}{\partial x^2_i} \ , \]
and $H_i$ be an operator on $\psi (x_i, t)$ defined as
\begin{equation}
H_i = -\frac{\hbar \sigma^2_i}{2} \nabla^2_i +  V_i(x_i) \ . 
\label{Hamiltonian}
\end{equation}
Then the equations~(\ref{update_rule11a}) can be rewritten as
\begin{equation}
\frac{\partial \psi_i (x, t)}{\partial t}
	= - \frac{1}{\hbar} H_i \psi_i(x_i, t) + \varepsilon_i (t) \psi_i(x_i, t)   \ . 
\label{update_rule11}
\end{equation}

When the differential equations~(\ref{update_rule11}) evolve into a stationary state (equilibrium), 
	they become
\begin{equation}
E_i \psi_i(x_i, t)  =  H_i \psi_i(x_i, t) , \quad \mbox{ for $i=1, 2, \ldots, n$} \ , 
\label{update_rule12}
\end{equation}
where $E_i$, $E_i = \hbar \varepsilon_i$, is a scalar.

For a physical system consisting of $n$ particles, 
	let $x_i$ be the position of particle $i$, $1 \le i \le n$, in the one dimensional space $\mathbb{R}$.
Let $\sigma^2_i = \hbar / m_i$,
where $m_i$ is the mass of particle $i$.
Then equations~(\ref{update_rule12}) become
\begin{equation}
E_i \psi_i(x_i, t)  =  \left( -\frac{\hbar^2}{2 m_i} \nabla^2_i + V_i(x_i) \right) \psi_i(x_i, t) \ . 
\label{stationary_state}
\end{equation}
They are the conditions for the physical system to be in a stationary state 
	when its dynamics is defined by the generalized APP algorithm.
Equation~(\ref{stationary_state}) is also the time-independent Schr\"{o}dinger equation.
(It is straightforward to generalize this derivation to three dimensions, but it does not yield any deeper understanding.)

In conclusion, from a pure mathematical observation,
	the time-independent Schr\"{o}dinger equation is derivable 
	from a soft-decision iterative decoding algorithm.
From the derivation we can see that 
	the soft decisions $\psi_i(x_i, t)$ of the decoding algorithm
	are the classic wavefunctions in the Schr\"{o}dinger equation.

\vspace{35\baselineskip}
~~~~~

%%\nocite{Schrodinger26a,Messiah99,Tegmark01}

%\bibliographystyle{../bib/IEEEtran}
%\bibliography{../bib/AIsfs}

\end{document}